\documentstyle[preprint,aps]{revtex}

\begin{document}
\draft
\preprint{}
\title{Dynamical Properties of a Growing Surface on a Random  Substrate}
\author{Dinko Cule }
\address{Department of Physics and Astronomy\\University of Rochester\\
Rochester, NY 14627}

\date{\today}
\maketitle
\begin{abstract}
 The dynamics of the discrete Gaussian model for the surface of a 
crystal deposited on a 
disordered substrate is investigated by Monte Carlo simulations.  
The mobility of the growing surface was studied as a
function of a small driving force $F$ and temperature $T$.
A  continuous transition is found from high-temperature phase
characterized by linear response to a low-temperature phase with
nonlinear, temperature dependent response. 
In the simulated regime of driving force the numerical
results are in general agreement with recent dynamic renormalization
group predictions.

\end{abstract}
\narrowtext

There has been considerable progress in 
recent investigations of crystalline surface growth \cite{R1,R2,R3}. 
It is known \cite{R4,R5} that, due to the 
discreteness and fluctuations (thermal fluctuations and fluctuations in the 
growth process itself), a crystalline surface undergoes a phase transition 
between a high-temperature rough phase and  a low-temperature smooth phase. The
presence of the quenched disorder in the crystal (either in the substrate
\cite{R6} or in the bulk \cite{R7}) changes both the critical temperature and 
the character of the low-temperature phase. Below critical temperature, $T_c$, 
the system develops a glassy phase characterized by the existence of many 
metastable states to which surface configurations are pinned 
by disorder. The surface itself remains rough but with very different static 
and dynamic properties \cite{R6,R7} compared to the roughness above $T_c$.

In the present work the dynamic properties of the roughening transition are
examined by numerical simulations. The numerical studies of the static
properties were reported elsewhere \cite{R8}. The simulated system is 
based on the Hamiltonian of the discrete Gaussian model which was very 
successful \cite{R4,R5} in describing the surface on a flat substrate:

\begin{equation}
 H = \frac{\kappa}{2}\sum_{\langle<i,j\rangle>}^{} (h_{i} - h_{j})^{2}.
\label{E1}
\end{equation}
\noindent
The sum runs over nearest-neighbor pairs, 
$\kappa $ is the surface tension,
and $h_i$ is the height of the surface above the point $i$ on the 
two-dimensional
basal lattice.  In the case of a flat surface $h_i$ takes integer 
values in the units of lattice spacing $a$ in the direction 
perpendicular to the surface.  To simulate the disordered substrate
a random quenched height $d_i$ chosen uniformly (and
independently for each site) in the interval $(-a/2,+a/2]$ was first
assigned to each site. 
The height $h_i$ then takes the values $h_i=d_i+n_{i}a$ where
$n_i$ is any  positive or negative  integer.
   
In the continuum limit, $h_i\rightarrow \phi(\vec{x})$, $d_i\rightarrow
d(\vec{x})$, and {\it near the critical point}, Eq. (\ref{E1}) maps to 
the Hamiltonian of the random phase sine-Gordon
model (RSGM) \cite{R4}:

\begin{equation}
 {\cal H} =  
      \int d\vec{x} \left\{   
                \frac{\kappa\beta}{2}\left[\nabla\phi(\vec{x}) \right]^{2}
 - g\cos\left( 2\pi\left[\phi(\vec{x}) - d(\vec{x})\right]/a\right)
\right\}.
\label{E2}
\end{equation}
\noindent
The periodic cosine term comes from the lattice discreteness and is
crucial for the existence of a phase transition. The constant $g$ might be
considered as  the
strength of the periodic pinning potential. The Hamiltonian (\ref{E2}) 
also describes other disordered systems: two-dimensional vortex glass with a
parallel magnetic field \cite{R9}, charge density waves, and it
is also equivalent to the vortex--free $XY$ model with random field.

  If the disorder is absent, the predicted value \cite{R1} for the
critical temperature of the discrete Gaussian model (\ref{E1}) 
is of the order
$T_R\approx 1.45$, which is close to the results from computer
simulations \cite{R10}. The behavior of the roughening transition in the
presence of an applied force $F$ was studied in detail by Nozi\`{e}res and
Gallet (NG)  \cite{R5}. They found a broadening of the transition 
due to nonequilibrium conditions with
crossover occurring below $T_R$. In the limit $F\rightarrow 0$, the
interface mobility $\mu = v/F$ has a sharp jump at $T_R$, from a finite
value for $T>T_R$ to zero for $T<T_R$. If surface is driven by a small,
but finite, driving force it remains rough 
(with temperature- and force-dependent mobility) 
even below $T_R$. The NG theory 
describing the system in the temperature region close to $T_R$ (``interrupted
renormalization'' scheme) or well below $T_R$ (homogeneous nucleation)
is in quantitative agreement with experiments on  $(0001)$ interfaces of
hcp $^4\mbox{He}$ crystals near the roughening transition \cite{R11}.

  The effects of the disordered substrate on the dynamic properties in
the vicinity of the phase transition were studied by the dynamic 
renormalization group (RG) methods
\cite{R6,R12} based on the Langevin dynamics with a small driving force $F$:

\begin{equation}
\frac{\partial \phi}{\partial t} = -\frac{\delta {\cal H}}{\delta \phi}
      + F + \xi.
\label{E3}
\end{equation}
\noindent
Here the Gaussian fluctuating noise  $\xi$ satisfies 
$\langle\xi(\vec{x},t)\xi(\vec{x}',0)\rangle =
2T\delta(\vec{x}-\vec{x}')\delta(t)$ 
and the exponentiated random phase $d(\vec{x})$ in Eq. (\ref{E2}) 
also obeys Gaussian  statistics:
$\langle\exp\{\imath 2\pi d(\vec{x})/a\} \exp\{-\imath 2\pi d(\vec{x}')/a\}
\rangle = a^2\delta(\vec{x}-\vec{x}')$. 
The theory predicts that above the critical
temperature, $T_c = \kappa/\pi$, the system responses linearly to the
applied
force, i.e., the mobility $\mu$ is finite and independent of $F$, 
while below $T_c$ the response
is nonlinear, characterized by the temperature-dependent exponent $\eta$:

\begin{equation}
\mu \sim \left\{
         \begin{array}{lr}
          (T/T_c -1)^{\zeta}  & \mbox{for $T> T_c$,} \\
          (1-T/T_c )^{\zeta}F^{\eta}  & \mbox{for $T< T_c$,} 
         \end{array}
         \right.
\label{E4}
\end{equation}
\noindent
where $\zeta \approx 1.78$ is a universal constant and 
$\eta=\zeta |1-T/T_c|$. 
At $T=T_c=\kappa/\pi$ and finite $F$, Eq. (\ref{E4}) is corrected by a term
$\sim |\ln F |^{-\zeta}$ so that mobility does not vanish \cite{R6}.
Also,  for $T<T_c$, Eq. (\ref{E4}) holds only if $F^{-|1-T/T_c|}\gg 1$.
If this condition does not hold, the original Eq. (48) in Ref.
\onlinecite{R6} has to be used.
According to RG analysis, these results are valid close to $T_c$ and in 
the large $L$, small $g$, and small $F$ limit with the crossover regime 
characterized by two effective
lengths: $L_g\sim g_{0}^{1/2|1-T/T_c|}$ 
and $L_F\sim F^{-1/2}$
($g_0=\pi g^2/2T^2$ is the bare coupling constant).

Numerical simulations of the model (\ref{E3}) with the Hamiltonian
(\ref{E2}) in the limit of small $g$  were recently performed by
Batrouni and Hwa \cite{R13} in the context of a randomly pinned planar
flux array. They found no sign of phase transition in statics but in
dynamics they observed a phase transition which is, however, only in
qualitative agreement with RG predictions (\ref{E4}). 
The constant $\zeta$ extracted from their data is significantly smaller
than that in Eq. (\ref{E4}).
On the other hand, the behavior of the surface under the influence of a
strong driving force  including the KPZ nonlinearity also has many 
interesting features and it has been the subject of recent
investigations \cite{R14}.

After a brief description of the simulated dynamics
we will present our numerical results for the
dynamics of the model (\ref{E1}) which maps to the model (\ref{E2}) with 
coupling constant $g$ of order 1.

 Every surface configuration can be completely specified by a
collection of column height variables $C=\{h_1,h_2, \ldots \}$. The
dynamics of the model is determined by
the transition rates $W(C\rightarrow C')$ 
which specify how the system evolves from a
given configuration $C$ into a new configuration $C'$. 
The probability $P(C,t)$ that the surface has configuration
$C$ at time $t$ is determined by the following master equation in
terms of these transition rates:

\begin{eqnarray}
\partial_t P(C,t)  &=& \sum_{C'}\{W(C'\rightarrow C) P(C',t) \nonumber
\\
                   & &  - W(C\rightarrow C') P(C,t) \}.
\label{E5}
\end{eqnarray}
\noindent

Without driving force, the system evolves to equilibrium and the transition
rates satisfy the detailed balance condition: $W(x)=W(-x)e^{-x}$,
where $x = \beta \Delta H$, and $\Delta H$ is change of energy. The
driving force $F$ is included by simply adding a term $\Delta n F$ to
$\Delta H$ in $W$, i.e., $W=W(\beta[\Delta H + \Delta n F])$ where
$\Delta n$ is a local change in the height
between configurations $C$ and $C'$. The commonly
used choice of $W$ is the Metropolis rate:

\begin{equation}
W(\beta[\Delta H  +\Delta n F]) = 
\mbox{min}\{1,e^{-\beta(\Delta H + \Delta n F)}\}.
\label{E6}
\end{equation}
\noindent

  The Monte Carlo (MC) simulations presented in this work were performed on 
the two-dimensional square lattice of
linear dimension $L=64$, and with periodic boundary conditions. The lattice
was divided into two sublattices. During the first half of the time step, all
heights $h_i$ of one sublattice were simultaneously updated by
increasing or decreasing them (independently) by one unit keeping the
heights of the other sublattice fixed. The moves are then accepted or
rejected according to the Metropolis rule (\ref{E6}) with Hamiltonian
(\ref{E1}) and constant $\kappa=2$. 
In the second half of the time step, the second sublattice is
upgraded keeping the first one fixed.

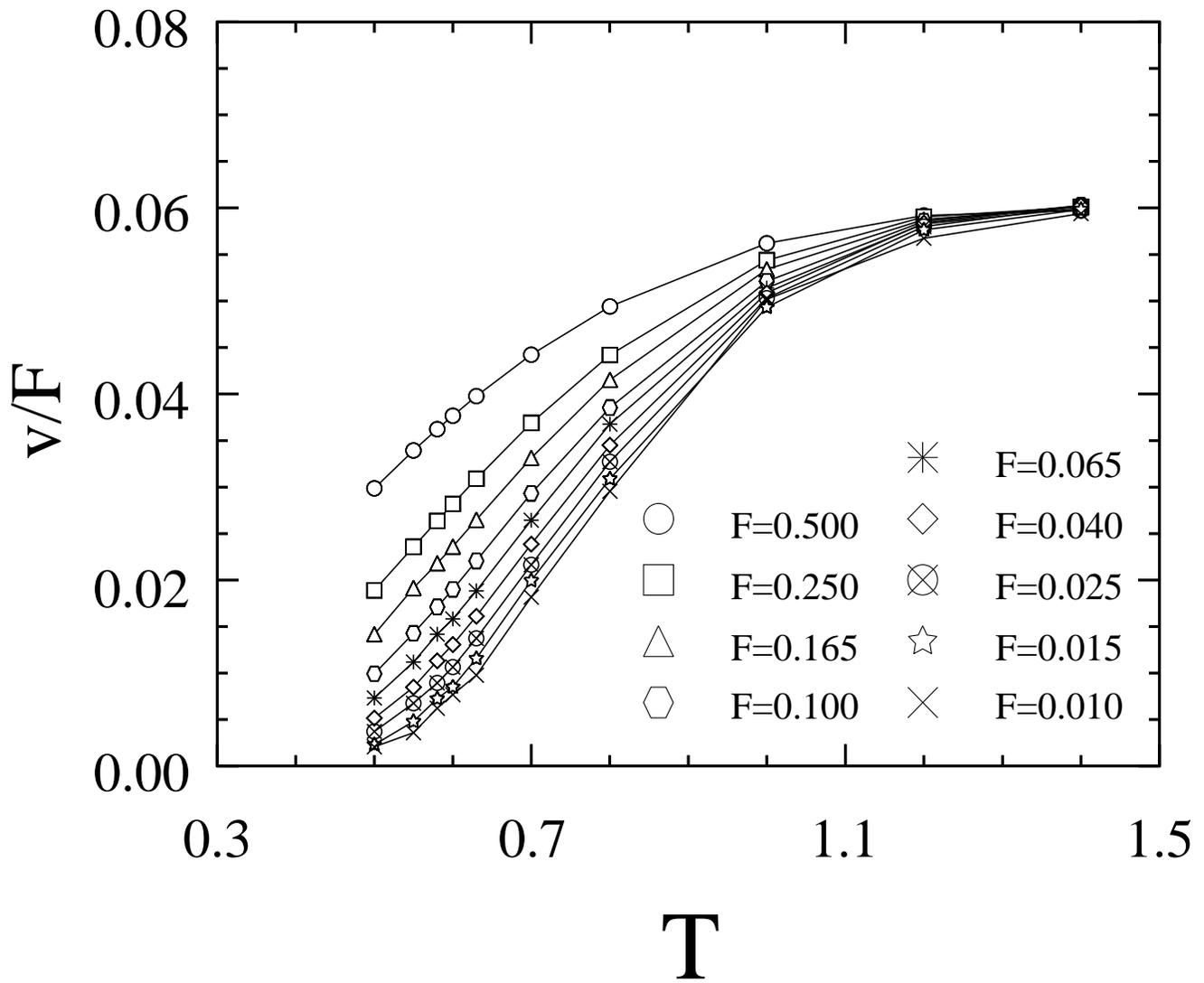
\begin{figure}
\caption{
The dependence of the mobility $\mu=v/F$ as function of temperature for
different $F$. The system size is $L=64$ and sample averaging is
performed over 50 realizations of disorder. The full curves are the
guides to the eye.
}
\label{F1}
\end{figure}

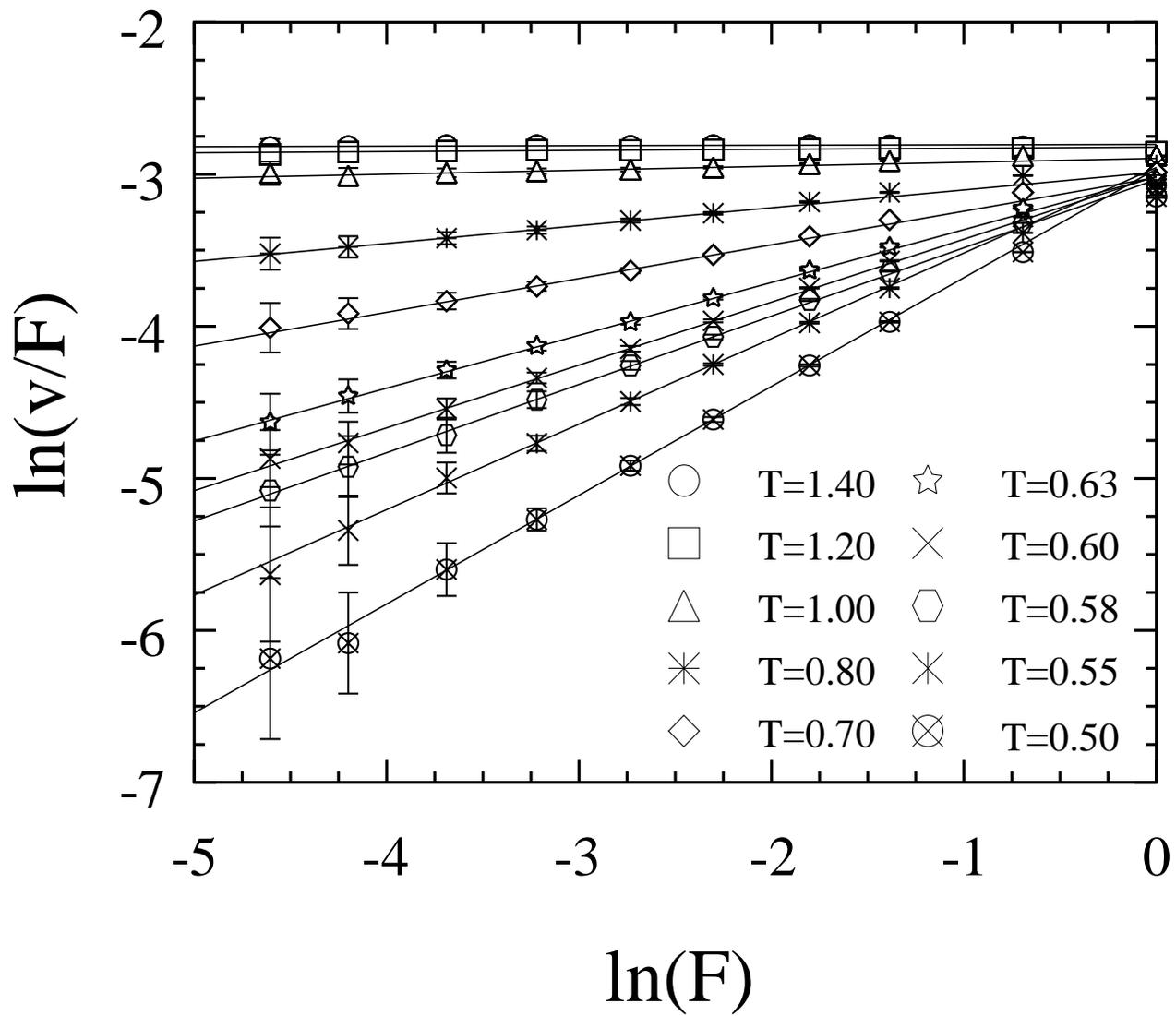
\begin{figure}
\caption{
The $\log$--$\log$ plot of the mobility versus driving force. The straight
lines are the best fits to $\ln(v/F) = a + b\ln(F)$ including
only six the smallest values of $F$ for each $T$ in Fig.
\protect\ref{F1}. 
}
\label{F2}
\end{figure}

  Starting with the equilibrated configurations saved after 
measuring the static properties of the system \cite{R8}, the force was
turned on by implementing (\ref{E6}). The velocity of the growing
surface averaged over different realizations of the disorder was monitored
as a function of MC steps. Typically, up to $10^4$ initial MC steps were
discarded since the  system requires some time to reach its stationary state
characterized by a uniform velocity. Measurements were performed over
additional MC steps whose lengths depended on the values of $F$ and $T$. 
The length of the runs ranged from $5\times 10^4$ 
(for large $F$ and  large $T$) to $10^6$ MC steps (for small $F$ and small $T$).
The average surface heights (in lattice units $a$)
at the end of the MC runs  were between several thousand steps for
large $F$ and $T$ to several dozen steps for small $F$ and $T$.
For every MC run, the measurements of the surface velocity were
grouped into several groups (usually about ten) and the average
velocity of these groups with corresponding error bars is presented
in Fig. \ref{F1} and Fig. \ref{F2}.
A practical problem in these simulations is to measure the response of the 
system to very small
force because very long MC runs are required in order to get reliable
statistics. We started with the forces of order 1 and then decreased
$F$  gradually toward the lowest value ( $F=0.01$) for which the data
analysis still suggests that the surface is moving with uniform velocity
although the error bars are much larger compared with the  measurements
with larger $F$ and $T$ (see Fig. \ref{F2}). For smaller values of $F$
we could not extract reliable  statistics  within  the  computer 
time available. 

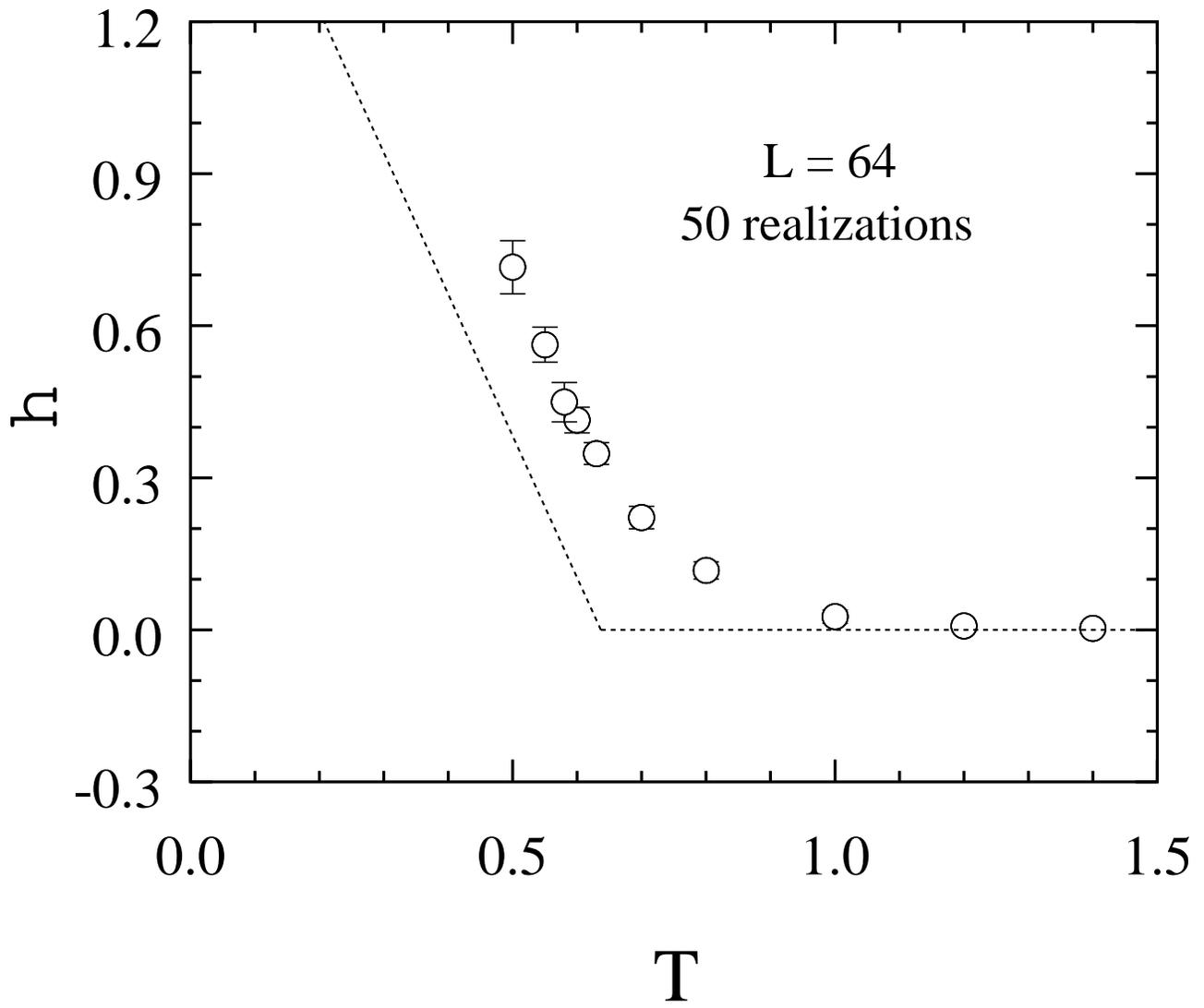
\begin{figure}
\caption{
Plot of the coefficient $b(T)$ (circles) from the fitting equation (see text and
Fig.  \protect\ref{F2}). The dotted line is the analytical prediction  
(\protect\ref{E4}). 
}
\label{F3}
\end{figure}

  Figure \ref{F1} shows the behavior of the mobility, $\mu = v/F$, as a 
function of
temperature for different driving forces, $F$, while Fig. \ref{F2}
shows the log-log plot of $\mu$ versus $F$ for different $T$. The
sample averages were
performed over 50 realizations of the disorder. 
It is clear from the figures
that the system has a phase transition from the regime at higher $T$,
where mobility is finite and independent of temperature and force, to the 
nonlinear regime at lower temperatures where $\mu$ depends on $F$ and
$T$. The transition itself is very broad and  the
position of the critical temperature $T_c$ is not very clear. 
The straight lines in Fig. \ref{F2}  are the best fits to the fitting equations 
$\ln(v/F)=a(T)+ b(T) \ln F$. 
Only the six lowest values of $F$ from Fig. \ref{F1} were
included in the fit: $F=0.010$, $0.015$, $0.025$, $0.040$, $0.065$, and $0.100$.
The slope of the lines, or coefficient $b$, corresponds to the exponent
$\eta$ in formula (\ref{E4}). Figure \ref{F3} shows a  comparison of the
exponent $\eta$ plotted according to Eq. (\ref{E4}) (dotted line)
and the corresponding values
extracted from the data in Fig. \ref{F2} (circles). 
Note that at $T=T_c$ Eq. (\ref{E4}) has to be corrected with the above
mentioned logarithmic contribution due to finiteness of $F$ so that 
disagreement between  the dotted line and  the numerical results is expected at
and very close to $T_c$.
Generally speaking, there
is an agreement between RG calculation and the numerical results.

   The finite size effects are examined by repeating the simulations for a
few temperatures below $T_c$ on the system size $L=128$ and with sample
averages over 25 realizations of the disorder. No significant
difference with respect to the $L=64$ results  was noticed. 
This is expected since, according to the RG analysis \cite{R6},
the size $L=64$ is already
larger than the effective crossover length: $L_F\approx 1/\sqrt{F} = 10$ 
(which is estimated using the smallest simulated value of $F$).

   To summarize, numerical simulations based on the Metropolis--type
dynamics for which the local detailed balance condition is always
satisfied were performed and compared with predictions of RG
calculations. The broad transition becomes sharper as applied force
becomes smaller. 
The broadening is a  consequence of 
the presence of the spatially varying pinning potential (due to
disorder)  with the strong coupling constant, 
the applied uniform external force, 
and (to a smaller degree) the finite system size.

  The comparison of the numerical results and RG predictions suggests
that there is a qualitative and to some degree even quantitative agreement
between the two, although numerical results suggest that the critical
temperature is shifted toward higher values in comparison to the
RG result.
The deviation from linear response occurs at temperatures larger than
$T_c$ of statics (the numerical value for $T_c$ of statics for model
(\ref{E1}) is $T_c=0.643\pm 0.006$ \cite{R8}). 
This is probably due to the strong coupling regime
where perturbative RG results are not expected to work. The
recent self-consistent, nonperturbative Hartree type calculations for 
relaxational
dynamics show that the critical temperature slowly increases with $g$
if $g$ is larger than some characteristic value below which $T_c$
does not depend on $g$ \cite{R15}. It is believed that this discrepancy
between critical temperatures in statics and dynamics is an effect of the
existence of many metastable states, and it is also found in
other physical systems \cite{R16}.

 I am  very grateful to Y. Shapir, J. Krug, and L.-T. Tang for most useful 
discussions and valuable comments. I acknowledge the kind hospitality of
the Institute for Theoretical Physics, Santa Barbara, where this work was
done.  I also thank Thinking Machines Corporation for use of their
computers. This research was supported by the National Science Foundation
under Grant No. PHY89-04035.

\end{document}